\documentclass[twocolumn,prc,showpacs,amsmath,amssymb,amsfonts]{revtex4}
\usepackage{bm}
\usepackage[final]{graphicx}

\newcommand{\beq}{\begin{equation}}
\newcommand{\eeq}{\end{equation}}
\newcommand{\beqa}{\begin{eqnarray}}
\newcommand{\eeqa}{\end{eqnarray}}

\newcommand{\bb}{\langle}
\newcommand{\kb}{\rangle}
\newcommand{\st}{\ensuremath{\sqrt{\sigma}}}
\newcommand{\rvec}{\mathbf{r}}
\newcommand{\bvec}[1]{\ensuremath{\mathbf{#1}}}
\newcommand{\bra}[1]{\ensuremath{\bb#1|}}
\newcommand{\ket}[1]{\ensuremath{|#1\kb}}

\newcommand{\bfalp}{\mbox{\boldmath{$\alpha$}}}

\newcommand{\half}{\frac{1}{2}}
\newcommand{\thalf}{\frac{3}{2}}
\newcommand{\shalf}{{\scriptstyle\frac{1}{2}}}
\newcommand{\sthalf}{{\scriptstyle\frac{3}{2}}}

\newcommand{\qmag}{|\bvec{q}|}

\newcommand{\ysa}[3]{\mathcal{Y}_{#1,#2}^{#3}}
\newcommand{\ysc}{\tilde{y}}

\begin{document}

\title{Virtual photon asymmetry for confined, interacting 
Dirac particles with spin symmetry}
\author{V.R. Pandharipande}
\email{vrp@uiuc.edu}
\affiliation{Department of Physics,
University of Illinois at Urbana-Champaign,
Urbana, Illinois 61801}
\author{M.W. Paris}
\email{mparis@jlab.org}
\affiliation{Thomas Jefferson National Accelerator Facility,
MS12H2, 12000 Jefferson Ave, Newport News, VA, 23606}
\author{I. Sick}
\email{Ingo.Sick@unibas.ch}
\affiliation{Departement f\"{u}r Physik und Astronomie,
Universit\"{a}t Basel, Switzerland}

\date{\today}

\begin{abstract}
We study the Bjorken $x$ dependence of the virtual photon spin asymmetry in
polarized deep inelastic scattering of electrons from hadrons. We use an
exactly solved relativistic potential model of the hadron, treating the
constituents as independent massless Dirac particles bound to an infinitely
massive force center. The potential is chosen to have spin symmetry and a
linear radial dependence with spherical symmetry. The effect of interactions of
the struck constituent with the remainder of the target on the longitudinal
photon asymmetry is demonstrated. In particular, the small-$x$ suppression of
the photon asymmetry observed in polarized deep inelastic scattering from the
proton is shown to be a consequence of these interactions. The effect of
$p$--wave components of the Dirac wave function, long known to give an
important contribution to the spin of hadrons, is explicitly demonstrated
through their interference with the $s$--wave term.
\end{abstract}
\pacs{13.60.Hb,12.39.Ki,12.39.Pn}

\maketitle

{\em Introduction.}--The spin dependent structure functions measured in deep
inelastic scattering of electrons from nucleons have recently been measured to
high precision \cite{Anthony:2002hy,Airapetian:2002rw,Zheng:2003un}.
Calculations of the polarized structure functions have been performed in an
array of models \cite{JaffeManohar90,Jaffe:1990qh,Thomas91,Song:1994ip,
Weigel:1996jh,Steffens:1998rw,Barik:1999ym,Gorchtein:2004}.

The spin symmetry \cite{Smith:1970,Bell:1975vq} and the closely related
pseudospin symmetry have been exploited to explain approximate degeneracies in
nuclear \cite{Ginocchio:1996zp} and hadronic \cite{Page:2000ij} spectroscopy.
In this communication we calculate the virtual photon longitudinal spin
asymmetry \cite{Filippone01} using a model with spin symmetry
\cite{Paris:2003at}.  The present work demonstrates the effect of including
interactions among the constituents of the composite hadronic target on
physical observables. This exact calculation allows us to make unambiguous
statements regarding the effects of interactions within the model and the
validity of commonly used approximations.

The model of Ref.\cite{Paris:2003at} is a quantum mechanical model for a single
massless Dirac particle confined by an linear potential, assumed valid to all
radii. It neglects the effects of $q\bar{q}$ pairs. The gluons are imagined to
have been integrated out resulting in the confining potential via a flux tube.
Ignoring radiative gluon corrections should not be an impediment in the
calculation of the spin asymmetry, observed in polarized electron scattering
experiments to be nearly independent of $Q^2$. The potential is assumed to be
one-half vector plus one-half scalar and therefore enjoys the spin symmetry
\cite{Page:2000ij}.  At small radii the potential is nearly zero and so, for a
given resonance, displays asymptotic freedom.

Previous calculations of inclusive unpolarized and polarized structure
functions performed within the cavity approximation to the bag model neglect
the role of a confining interaction. The naive bag model assumes that the
constituents of the hadron are free and on mass-shell; confinement only
restricts the momentum of the constituents. It is interesting to study the
consequences of a confining interaction which takes the constituent off the
mass-shell in an exactly solvable model.

The present calculation is similar to the bag model calculations in
Refs.\cite{Jaffe:1975nj,Thomas91,JaffeJi92,Song:1994ip}.  These works treat the
struck constituent as a free particle whose state is described by a plane wave.
Here we use an exactly solved single particle relativistic potential model of
the hadron. The eigenstates of this Hamiltonian, which are four-component Dirac
spinors, describe the state of the struck constituent. Our model calculation is
exact and includes eigenstates with a maximum excitation energy of about 10 GeV
\cite{Paris:2003at}.

We compare the exact result to the plane wave impulse approximation (PWIA) and
find good agreement for the longitudinal photon asymmetry $A^q_1(\xi)$ to the
level of about ten percent. The qualitative agreement of the exact calculation
with PWIA allows us to compare to a bag model PWIA evaluation of the asymmetry. 
We attribute differences between the asymmetries calculated in the present
model and the bag model to interactions.

Moreover, the present work shows the effect that interactions have on the spin
structure of hadrons within a relativistic potential formalism. It has long
been known that $p-$wave components, which are necessary in the Dirac
description of confined particles, reduces the contribution of valence quarks
to the spin of the nucleon \cite{Bogoliubov:1968,Kuti71,Close:1974ux,
Chodos:1974pn,Jaffe:1975nj,JaffeJi92,JaffeManohar90}.  We emphasize the role of
the (lower component) $p-$wave terms of the ground state Dirac wave function in
determining the $x$ dependence of the $A^q_1(\xi)$. The interference between
the dominant $s-$wave and the $p-$wave parts of the valence quark Dirac wave
function suppresses $A^q_1(\xi)$ at small values of $\xi$. This novel
observation demonstrates unambiguously an effect of treating hadronic
constituents as bound Dirac particles. It suggests that interactions among
the constituents reduce the contribution of valence quarks to the spin of
nucleons.

{\em Model calculation.}--We consider the calculation of the virtual photon 
spin asymmetry in DIS of a charged leptonic probe from a hadronic target 
within the model of Ref.\cite{Paris:2003at}. The model Hamiltonian is 
chosen as 
\beq
\label{eqn:H}
H=\bfalp\cdot\bvec{p} + \frac{1+\beta}{2}\st r,
\eeq
where $\bfalp$ and $\beta$ are Dirac matrices in the standard representation
\cite{BjD}.  It describes a massless Dirac particle in a linear confining well.
The half-vector plus half-scalar structure of the confining potential is chosen
for its spin symmetry \cite{Page:2000ij} wherein spin-orbit doublets are
degenerate. Relatively small spin-orbit splittings seen in meson spectra
motivate this choice. Computations are simple with this choice since the lower
components of the wave function are not coupled by the potential.  The value of
the string tension $\st$ is assumed to be 1 GeV/fm as indicated by the slopes
of baryon Regge trajectories.  In Ref.\cite{Paris:2003at} all the eigenstates
of this model were obtained exactly for excitation energies up to $\sim 12$
GeV.  The ground state energy, $E_0$ for this string tension is 840 MeV.  The
model may be viewed as a heavy-light meson, such as $\bar{t}u$, in the limit
that the antiquark mass goes to infinity.  However, it retains only the
confining part of the $\bar{t}u$ interaction modeled by a flux tube. 

The model neglects gluon and sea-quark contributions to DIS as well as the QCD
evolution. However, the observed ratio of the spin structure function $g_1(x)$
to $F_1(x)$, the unpolarized structure function, is relatively independent of
$Q^2$ \cite{Filippone01}. Our objective is to calculate the $x$-dependence of
this ratio for the contribution of valence quarks to DIS and we hope that the
model is useful in this context. The utility of our potential model is limited,
however, and we note that it has known shortcomings. It cannot, for example, 
reproduce the observed ratio of $g_1^p/g_A$ \cite{Abbas:1989hg}.

The virtual photon asymmetry is defined as \cite{Filippone01}
\beq
\label{eqn:A1}
A_1=\frac{\sigma_{\shalf}-\sigma_{\sthalf}}{\sigma_{\shalf}+\sigma_{\sthalf}}
\eeq
with $\sigma_{\shalf}$ and $\sigma_{\sthalf}$ the helicity cross sections
for the target angular momentum antiparallel and parallel to the photon
helicity, respectively. We may calculate the inclusive virtual 
photon helicity cross sections in the rest frame of the target as
\beqa
\label{eqn:sig1/2}
\sigma_{\half} &=& \sigma_M \sum_{I}
\left|\bra{I}\alpha_+ e^{i\qmag z}\ket{0,-{\shalf}}\right|^2
\delta(E_I-E_0-\nu) \\
\label{eqn:sig3/2}
\sigma_{\thalf} &=& \sigma_M \sum_{I}
\left|\bra{I}\alpha_+ e^{i\qmag z}\ket{0,+{\shalf}}\right|^2
\delta(E_I-E_0-\nu)
\eeqa
where $\qmag$ and $\nu$ are the momentum and energy transferred to the 
target, $\sigma_M$ is the Mott cross section and we assume that the
virtual photon is in the $\hat{\bvec{z}}$ direction. The ground states 
$\ket{0,j_z={\scriptstyle\pm\shalf}}$ have the total angular momentum 
projection $j_z=\pm\half$. The operator $\alpha_+$ corresponds 
to a virtual photon with positive helicity, and $\ket{I}$ are 
eigenstates of the Hamiltonian $H$ [Eq.(\ref{eqn:H})] with energies $E_I$.  

\renewcommand{\bottomfraction}{0.95}
\begin{figure}[b]
\includegraphics[ width=230pt, keepaspectratio, angle=0, clip ]{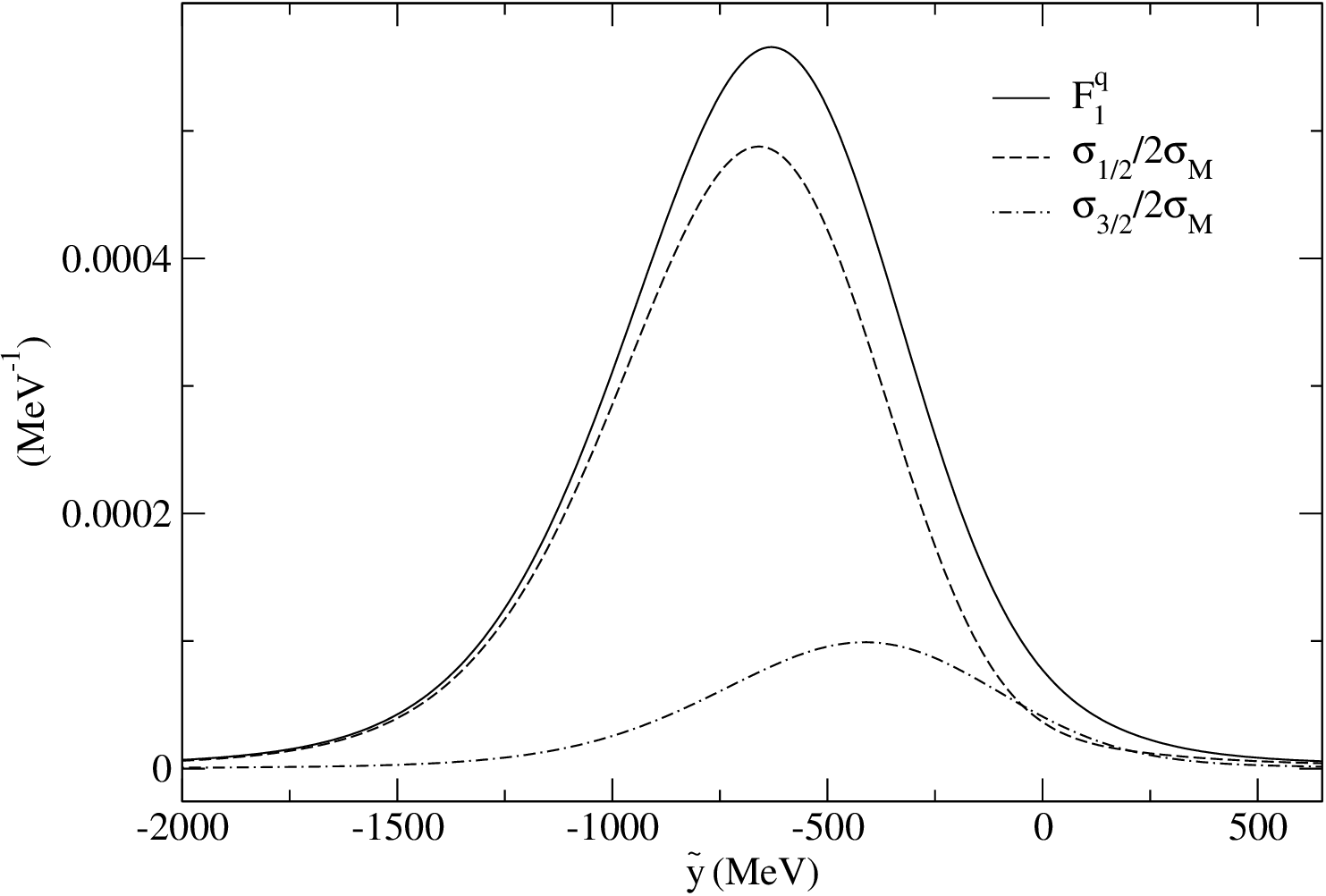}
\caption{\label{fig:hxs} Virtual photon helicity cross section of a confined 
massless quark, modulo twice the Mott cross section, as a function of $\ysc$.
The dashed $(\sigma_{\shalf})$ and dash-dotted $(\sigma_{\sthalf})$ curves
sum to the unpolarized structure function (solid curve).}
\end{figure}

The calculation of the virtual photon helicity cross sections
proceeds in this model, without approximations, exactly as the calculation 
of the unpolarized structure
functions described in Ref.\cite{Paris:2003at}. When $\qmag$ is large 
the $\sigma/\sigma_M$ depend only on $\ysc=\qmag-\nu$.  Figure (\ref{fig:hxs})
shows the calculated $\sigma_{\shalf}/(2\sigma_M)$ 
and $\sigma_{\sthalf}/(2\sigma_M)$
plotted as a function of the scaling variable $\ysc$, 
and their sum
\beq
\label{eqn:F1q}
F^q_1(\ysc) = 
\frac{1}{2\sigma_M}\left(\sigma_{\shalf}+\sigma_{\sthalf}\right),
\eeq
the unpolarized structure function.
The conventionally defined Bjorken and Nachtmann scaling variables 
are related to $\ysc$ by \cite{BPS00}: 
\beq
x (Q^2 \rightarrow \infty) = \xi = -\frac{\ysc}{M_T},
\label{eq:xyz}
\eeq
where $M_T$ is the target mass.
Thus small (large) negative $\ysc$ correspond to small (large) $x$. 
We note that the $\sigma_{\shalf}(\ysc)$ and $\sigma_{\sthalf}(\ysc)$ are not 
proportional, which implies that the $A_1^q$ of a confined relativistic quark 
has a large $\ysc$ or equivalently $x$ dependence. 

{\em Results.}--The ground state $|0,j_z\rangle$ of the confined quark 
has wave function:
\beqa
\label{eqn:gswf}
\Psi_{0,j_z}(\rvec) &=& \left(
\begin{array}{c}
f_0(r)\ysa{\half}{j_z}{0}(\bvec{\hat{r}}) \\
ig_0(r)\ysa{\half}{j_z}{1}(\bvec{\hat{r}}) \end{array}\right),
\eeqa
where $f_0(r)$ and $g_0(r)$ are the radial functions for the $s$--
and $p$--waves, respectively, and 
$\ysa{j}{j_z}{\ell}$ are the spin-angle functions obtained by 
coupling spin--$\half$ and orbital angular momentum $\ell$ to $j=\half$. 

\begin{figure}[b]
\includegraphics[ width=230pt, keepaspectratio, angle=0, clip ]{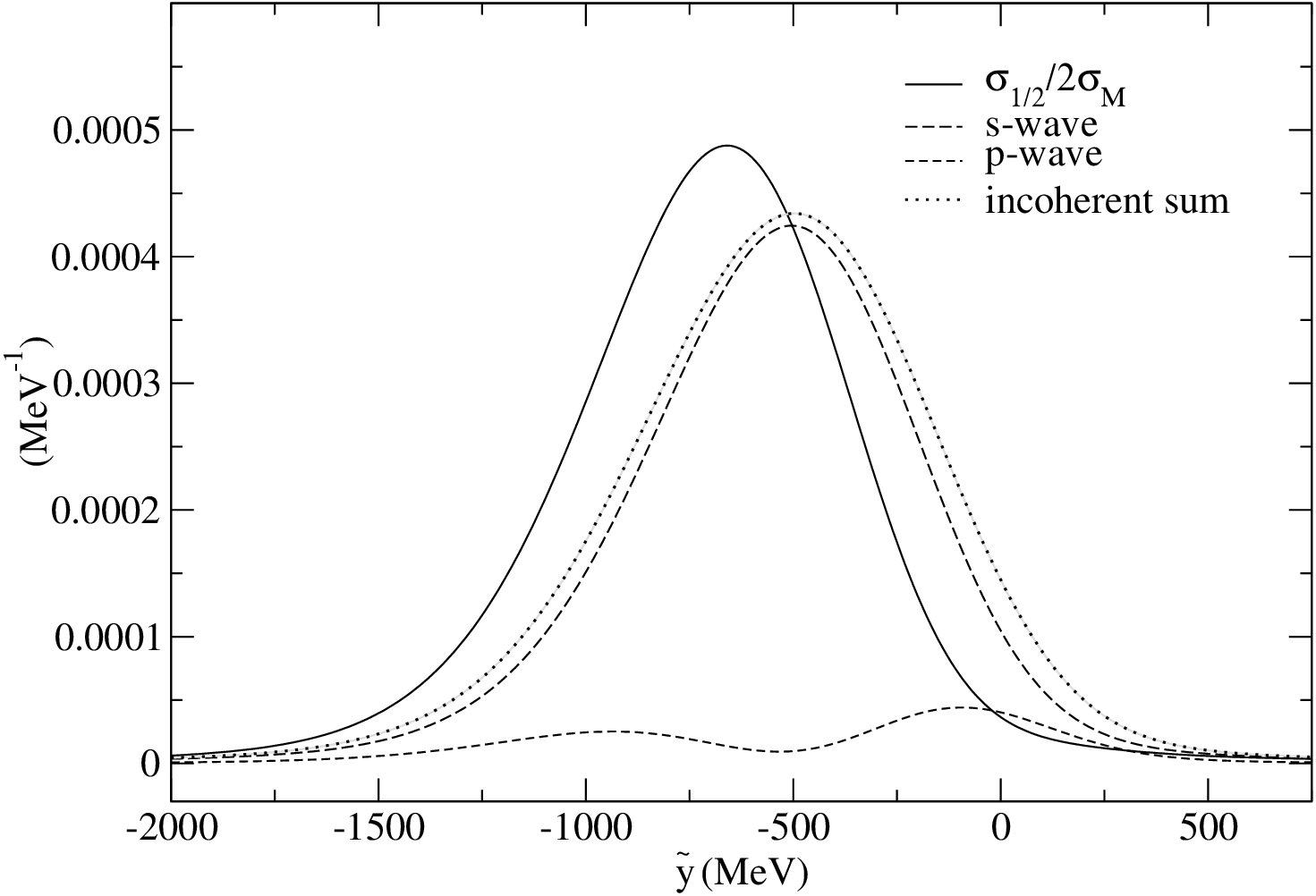}
\caption{\label{fig:intfr} Interference effects in $j_z=-\half$
($\sigma_\half$) structure function. The dashed lines give the contributions 
of the $s$-- and $p$--waves alone, the dotted line shows their incoherent sum 
and full line is the exact result. }
\end{figure}

The interference in the DIS between the $s$-- and $p$--waves 
contributes significantly to the $\ysc$ dependence of the 
$\sigma_{\shalf}$ helicity cross-section, $A_1^q$ and $F_1^q$. The effect
of interference is shown in Fig.(\ref{fig:intfr}) where we compare the 
polarized cross section $\sigma_\half$ including interference terms (solid
curve, labeled `full') with the polarized cross section neglecting 
interference terms (dotted curve). Also shown are the polarized cross 
sections obtained after keeping only the $s$-- or $p$--waves in the  
$j_z=-\half$ target. We note that the interference shifts 
$\sigma_\half$ to more 
negative $\ysc$ corresponding to larger values of $\xi$.  
Only the $p$--waves contribute to $\sigma_{\sthalf}$ shown 
in Fig.(\ref{fig:hxs}). 

The virtual photon asymmetry is given in terms of the spin-dependent
structure functions $g_1$ and $g_2$ \cite{Filippone01} by
\beq
\label{eqn:A1g1}
A_1=\frac{g_1 - \gamma^2 g_2}{F_1} \approx \frac{g_1}{F_1}
\eeq
where $\gamma^2=4M_T^2 x^2/Q^2$, in the scaling regime,
$Q^2\rightarrow\infty$.  As mentioned earlier, the observed 
$A_1$ of the proton, $A_1^p$ is nearly independent of $Q^2$, 
and is used to extract values of $g_1^p/F_1^p$ \cite{Filippone01}. 

Using the structure functions given in Fig.(\ref{fig:hxs})
we can easily calculate 
the virtual photon asymmetry $A^q_1$ or equivalently the ratio 
$g^q_1/F^q_1$ for a single confined quark, as a function of $\ysc$.  
In order to compare it with the data on protons we have to convert 
it to a function of $\xi$ by providing a mass scale $M_T$ (see 
Eq.(\ref{eq:xyz})).  Our model target has infinite mass associated with 
the center of the confining potential.  However, that mass is 
not relevant since only the confined quark contributes to DIS.  We 
use $M_T=2.5$ GeV $\sim 3E_0$, where $E_0$ is the energy of a single 
confined quark in the ground state.  With this choice the $F_1^q(\xi)$ 
becomes small at $\xi \sim 0.8$ as in the proton. The fact that the 
model target has infinite mass means that response can be non-zero, in 
principle, at arbitrarily large values of $\xi>0$. In fact the
calculated structure functions shown in Figs.(\ref{fig:hxs}) and
(\ref{fig:intfr}) are very close to zero at $\ysc=-2000$ MeV 
corresponding to $\xi\approx 0.8$. Nevertheless, the present model 
should not be used for values of $\xi\gtrsim 0.8$.

\begin{figure}
\includegraphics[ width=220pt, keepaspectratio, angle=0, clip ]{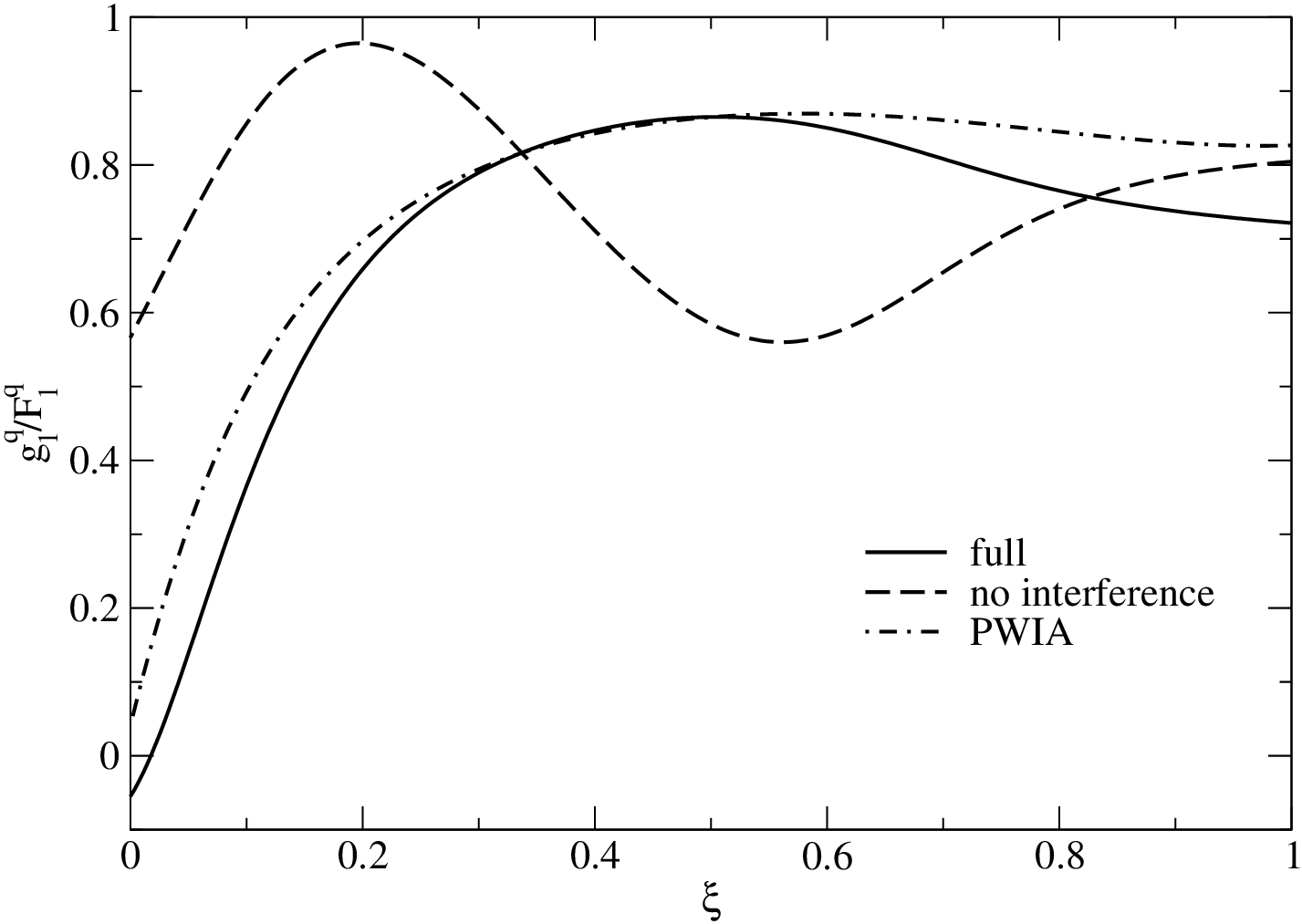}
\caption{\label{fig:g1of1} The $g^q_1/F^q_1$ for a single massless quark
confined by a flux-tube, as a function of the Nachtmann $\xi=(\qmag-\nu)/M_T$
with and without interference terms (see text) and compared to PWIA. The
curves are valid for $\xi\lesssim 0.8$.}
\end{figure}

The solid curve in Fig.(\ref{fig:g1of1}) 
shows  the $A_1^q(\xi)$ or equivalently $g_1^q(\xi)/F_1^q(\xi)$ of a 
confined quark. The calculated ratio goes to zero at small $\xi$, and 
this behavior is independent of the chosen value of $M_T$.  The dip 
at $\xi=0$ is due to the shift of $\sigma_\half$ to larger values of 
$\xi$, produced by the interference effect shown in Fig.(\ref{fig:intfr}). 
When the interference 
terms are omitted we obtain the dashed curve in Fig.(\ref{fig:g1of1}) 
which has $g_1^q/F_1^q \sim 0.6$ at $\xi=0$.  

Alternatively we could have chosen the string tension $\sqrt{\sigma}$ 
such that $3E_0=M_N$, the nucleon mass.  However, since $\sqrt{\sigma}$ 
provides the only mass scale in the Hamiltonian $H$ [Eq.(\ref{eqn:H})], 
$A_1^q(\xi)$ is independent of this choice.

Let's compare the exact calculation of $A^q_1(\xi)$ to that obtained
in plane wave impulse approximation (PWIA). 
We replace the final state $\bra{I}$ in Eqs.(\ref{eqn:sig1/2},
\ref{eqn:sig3/2}) with a positive energy plane wave 
$\bra{u_{\bvec{k}+\bvec{q},s}}$ with momentum $\bvec{k}+\bvec{q}$
and spin projection $s$ and replace the energy of the struck constituent
by that of a free particle: $E_I\rightarrow |\bvec{k}+\bvec{q}|
+ \langle V \rangle_0$; here $\langle V \rangle_0$ is the expectation
value of the potential in the ground state, chosen to reproduce the
first moment of the exact result. We may simplify the resulting expression 
for $\Delta\sigma=\sigma_{\half}-\sigma_{\frac{3}{2}}$ in PWIA using the
Dirac equation,
\beqa
\label{eqn:DEu}
f'_0(r) &=& E_0 g_0(r) \\
\label{eqn:DEl}
g'_0(r) + \frac{2}{r}g_0(r) &=& -\left[ E_0 - \st r \right] f_0(r)
\eeqa
for the ground state. Note the simplicity of Eq.(\ref{eqn:DEu}) owing
to the form of the Dirac structure of the potential in Eq.(\ref{eqn:H})
required by spin symmetry. In PWIA we obtain
\beqa
\label{eqn:dspwia}
\Delta\sigma &=& \frac{\sigma_M}{2}
\left\{
\left[\half\left(1+\frac{4\ysc^2}{E_0^2}\right)\right]
\int_0^\infty dk_\perp k_\perp \left|\tilde{f}_0(\bar{k})\right|^2 \right.
\nonumber\\
&-& \left. \frac{1}{E_0^2}\int_0^\infty dk_\perp k_\perp \bar{k}^2
\left|\tilde{f}_0(\bar{k})\right|^2 \right\},
\eeqa
where $\bar{k}=\sqrt{k_\perp^2+(E_0/2+\ysc)^2}$ and we have used the
fact that $\langle V \rangle_0 = E_0/2$ \cite{Paris:2003at}.
The PWIA for $A_1^q(\xi)$ is shown as a dot-dashed curve in 
Fig.(\ref{fig:g1of1}). The final state interactions (FSI) do not
change the qualitative behavior of $A_1^q(\xi)$ in the present model,
however they contribute to its suppression at small $\xi$.

\begin{figure}
\includegraphics[ width=200pt, keepaspectratio, angle=0, clip ]{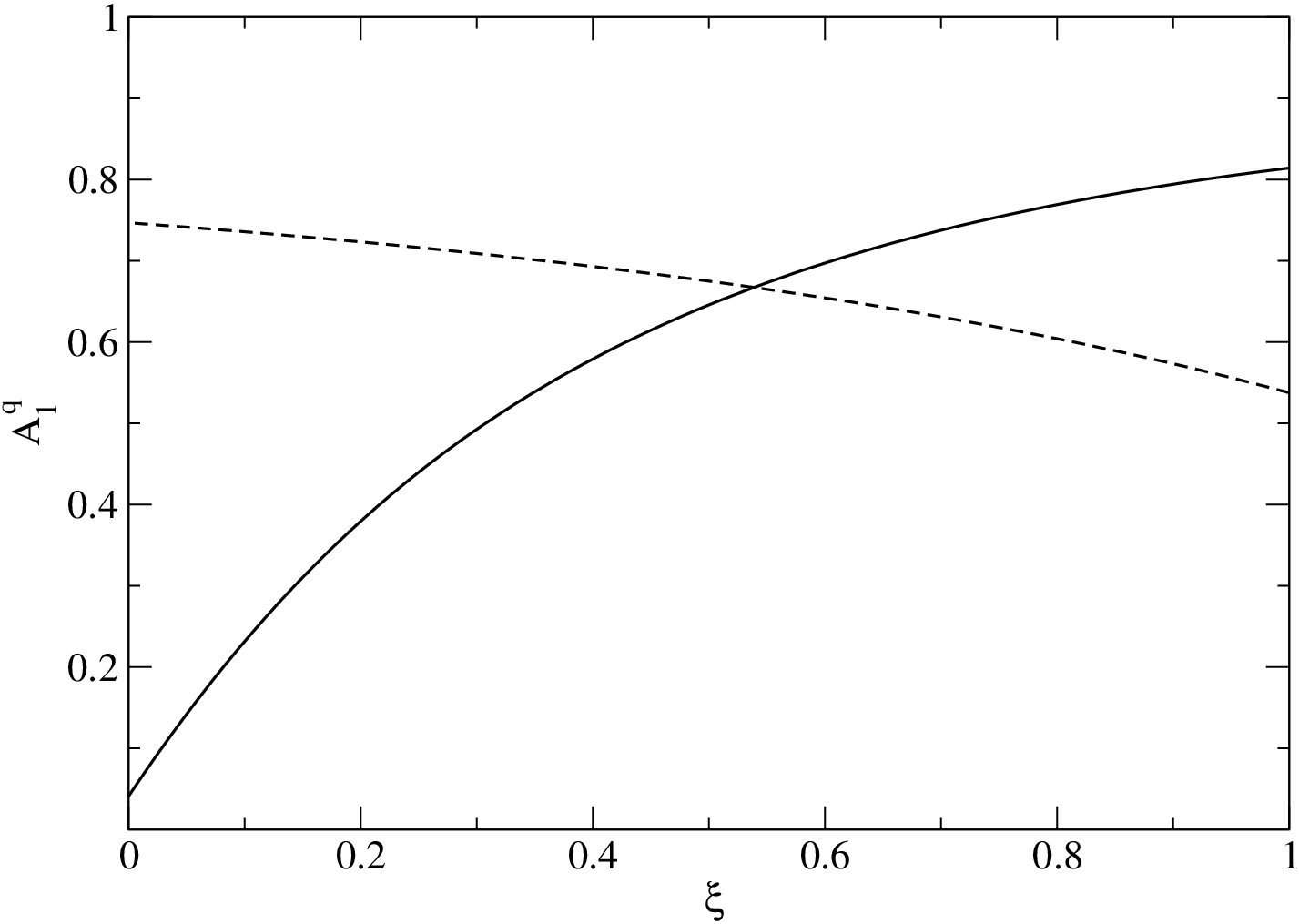}
\caption{\label{fig:a1qcfbm} The $A^q_1$ in the linear confinement model 
(solid curve) and in the cavity approximation to the bag model (dashed 
curve) versus the dimensionless variable $\xi$ described in the text.}
\end{figure}

We compare the results of the present model with those of the cavity
approximation to the bag model \cite{Chodos:1974je,Chodos:1974pn}.  We neglect
FSI in the bag model and compare the results for $A_1^q$ within PWIA. The bag
model wave function has the same form as Eq.(\ref{eqn:gswf}) with the radial
functions, $f_0(r)=n_0 j_0(pr)$ and $g_0(r)=n_0 j_1(pr)$ where $n_0$ is a
normalization factor and $j_\ell$ are the spherical Bessel functions. The value
of $p$, the momentum of the confined constituent in a cavity of radius $R$ is
fixed by a boundary condition and has the numerical value $pR\approx 2.04$.
We choose the cavity radius $R=0.65$ fm, the {\em rms} radius of the ground
state in the present model. The results are shown in Fig.(\ref{fig:a1qcfbm})
where the asymmetries for the linear confining model and the bag model are
plotted versus $\xi = -\ysc/E_0$, where $E_0$ is the ground state energy of the
constituent and takes on the value 0.84 GeV in the linear confinement model and
0.62 GeV in the bag model. The photon point $\xi=0$ is, of course, scale
invariant and independent of whatever mass scale one uses to obtain a
dimensionless $\xi$.

The experimentally observed suppression of the spin asymmetry in the proton at
small values of Bjorken $x$ (or $\xi$) is seen in the linear confinement model
but not in the bag model. Note that, although the bag model has similar
interference terms of the upper ($s$--wave) and lower ($p$--wave) terms of the
wave function these terms do not lead to a suppression at small values of $\xi$
of $A^q_1(\xi)$. The present model demonstrates that the $p$--waves give rise
to a dynamical suppression of the helicity distribution at small $x$ when
interactions are taken into account. It is, of course, possible to obtain the
suppression at small $\xi$ in the bag model by taking into account spin and
flavor dependent quark interactions \cite{CT88}. We obtain this 
suppression naturally as a consequence of the Dirac character of the
interacting constituent.

{\em Conclusion.}--In conclusion, the present work suggests that the 
$\xi$ dependence of 
$A_1^p(\xi)$ is a consequence of the interactions among the relativistic
fermionic constituents. The $p$--waves in bound quark wave functions 
interfere with the dominant $s$--waves to suppress $A_1^q(\xi)$ at small 
$\xi$ when the flux tube model for confinement is used. Although these 
interference terms are also present in the bag model they do not lead to 
a suppression of $A_1^q$ at small $\xi$.

Our model is certainly too simple; it approximates the 
problem of three interacting quarks by a relativistic one-quark 
problem.  Nevertheless $p$--waves occur very naturally in the 
wave functions of spin-half relativistic particles, and their effect 
will presumably exist in more refined treatments of spin asymmetries. 

This work has been supported by the US National Science Foundation 
via grant PHY-00-98353, and by the US Department of Energy 
under contract W-7405-ENG-36 and the Schweizerische Nationalfonds.
 
\bibliographystyle{apsrev}
\bibliography{psf}

\end{document}